\documentclass[11pt]{article}
\usepackage{amsmath} 
\usepackage{units} 
\usepackage{amsfonts} \parindent 0pt \parskip.2cm \topmargin -1.0cm
\oddsidemargin=0.25cm\evensidemargin=0.25cm \newfont{\bbbold}{msbm10
  scaled \magstep1}

 \newcommand{\D}{{\Delta}} 
 \newcommand{\m}{{\mu}}

\def\pslash#1{{\setbox0=\hbox{$#1$}
    \rlap{\ifdim\wd0>.7em\kern.22\wd0\else\kern.1\wd0\fi /}#1}}

\def\be{\begin{equation}} \def\ee{\end{equation}}
\def\ba{\begin{array}} \def\ea{\end{array}}

\begin{document}
\pagestyle{empty}
\begin{flushright}
\end{flushright}

\begin{center}
{\Large\sf Conjugate variables in quantum field theory: the basic case }\\[10pt]
\vspace{.5 cm}

{Klaus Sibold${}^1$ , Gautier Solard${}^2$}
\vspace{1.0ex}
 
\end{center}

{\small
\noindent                      
${}^1$Institut f\"ur Theoretische Physik, Universit\"at Leipzig,
Postfach 100920, D-04009 Leipzig, Germany\\  
${}^2$\'Ecole Normale Sup\'erieure, D\'epartement de Physique, Paris, 
France\\
}


\vspace{2.0ex}
\begin{center}
{\bf Abstract}
\end{center}

\noindent
Within standard quantum field theory of one scalar field we define
operators conjugate to the energy-momentum operators of the theory.
They are singled out by calculational simplicity in Fock space. In terms of
the underlying scalar field they are non-local. We establish their algebra
where it turns out that time and space operators do not commute.
Their transformation properties with respect to the conformal
group are derived. Solving their eigenvalue problem permits to reconstruct
the Fock space in terms of the eigenstates. It is indicated how Paulis
theorem may be circumvented. As an 
application we form the analogue of S-matrices which yields information
on the structure of the underlying spacetime. Similarly we define fields
and look at their equations of motion.\\

\newpage
\pagestyle{plain}
\section{Introduction}
``Space, time, matter'' has always been a great theme in theoretical physics
\cite{Weyl}. Whereas for Hermann Weyl in 1918 this meant to explain what general 
relativity says to this subject it is by now the task to reconcile quantum mechanics
and general relativity, a goal from which we seem to be far away. Special relativity and quantum mechanics, however seem to live well with each other as being represented in 
the form of relativistic quantum field theory. So is, 
e.g.\ modern particle theory at present  well described by a quantum field 
theory over flat Minkowski spacetime. And, more specifically, it is quite 
remarkable that, within the standard model of strong, electromagnetic and 
weak interactions, {\sl perturbative} considerations based on the Fock space 
of free fields are successful to a truly astonishing degree \cite{Kraus},
\cite{HeinmeyerHollikWeiglein}. Effects of curved background,
of non-commutative spacetime or
any other generalization of geometry are definitely small and may,
hopefully, also admit a kind of perturbative treatment. In this spirit it
has been proposed \cite{Sibold}  to study conjugate
variables in quantum field theory and thus to provide building blocks
for symplectic structures in quantum field theory.
Roughly speaking the construction runs along the following lines. In a first
step one defines pre-conjugate operators $X_\nu$ which satisfy commutation
relations of the type 
\be\label{precoop}
\left[P_\mu, X_\nu\right] = i\eta_{\mu\nu}O,
\ee
where the $P_{\mu}$ are the energy-momentum operators of the model and $O$
represents an operator 
through which one can ``divide'' in some sense, such that in a second step
one may obtain
\begin{equation}\label{cooop} 
\left[P_\mu, Q_\nu\right] = i\eta_{\mu\nu},
\end{equation}
for $Q_{\nu}$ being eventually the looked for coordinate operators. In fact,
it will turn out that we use only
\begin{equation}\label{precooopfin} 
\left[P_\mu, Q_\mu\right] = i\eta_{\mu\mu},\quad{\hbox{no sum}}
\end{equation}
for defining $Q_{\mu}$, where then the complete algebra of the $P_{\mu}$ and $Q_{\nu}$ 
constitutes the resulting symplectic structure.\\
A general remark concerning our treatment of the operators involved 
is in order here. We do not study their domains, but tacitly assume that
those would permit our calculations. If already on this formal level we
were to meet obstacles then we had just to stop the analysis. If, however
we succeed then still quite some work is ahead of us, an effort which we do
not undertake at present. In this context
we have to mention what is called Paulis theorem \cite{Pauli}: taken at
face value it would forbid the construction of a self-adjoint time operator
conjugate to the Hamiltonian, because a time coordinate would have to run
through all real numbers whereas the energy is restricted to be
non-negative (more precisely: bounded below). We shall however see
(down in sect.\ 4.1) how we hopefully circumvent the theorem.\\
The choice of $X_{\mu}$ is guided by qualitative considerations. In the present
paper they are selected as bilinear products of $a^\dagger, a$ -- creation and annihilation operators of a free scalar field -- where we demand only calculational 
simplicity in the aim to realize algebras, to solve eigenvalue
problems and the like. Hence we impose as few constraints as possible, still
permitting the construction of $Q_\mu$ as tools which can be handled on a
technical level.\\
In a companion paper \cite{SiboldEden} we put, to the contrary, as 
many constraints as to make the pre-conjugate operators unique: we realize 
the $X_{\mu}$ as {\sl charges} associated with a symmetry, hence demand that 
they can be formulated 
as local operators, bilinear in terms of fields, maintaining at the same time
Lorentz covariance. The hope is that such operators admit extension to all orders of perturbation theory. They turn out then to be the generators of conformal 
symmetry. We therefore call this the ``conformal case'', whereas we refer to
the first version, to be studied in the sequel, as the ``basic case''. (These
remarks explain the title.)\\
The present paper is self-contained; the reading of \cite{Sibold} or
\cite{SiboldEden}  is not required. It is organized as follows. In sect.\ 1 we define the $X_\mu$ and $Q_\mu$
and find in this context a new characterization of the dilatation operator $D$
and the conformal generator $K_0$. The operator $O$ of (\ref{precoop}) turns out to
be the number operator $N$ of Fock space. In sect.\ 2 we study the algebraic
properties of $X_\mu$ and $Q_\mu$, in particular their behaviour under Lorentz and conformal transformations. (Here we correct some calculational mistakes which
occurred in \cite{Sibold}.) In sect.\ 3 we solve the respective eigenvalue problems,
discuss normalization, completeness and basis independence and reformulate the
original Fock space in terms of eigenstates of $X_\mu$ resp.\ $Q_\mu$. Sect.\ 4
is devoted to applications: first we construct $S$-matrices and discuss how they
are related to an underlying spacetime; then we define fields as functions of $Q_\mu$ and look at their equations of motions. Finally, in sect.\ 5 we 
summarize our results and draw some further conclusions. In the appendix
we collect some formulae as to make the paper sufficiently self-contained.\\

\section{Definition of $X_\mu$ and $Q_\mu$}

As suggested in \cite{Sibold} we search as candidates for $X_\mu$ amongst Hermitian bilinear
products of $a^\dagger$ and $a$ (in this order), multiplied with factors $p_{\mu}$
and derivatives thereof, hence in $x$-space amongst bilinear Wick products of fields  $\phi(x)$, factors $x$ and derivatives with respect to $x$, integrated over three-space, if (!) the operators 
written in terms of creation and annihilation operators give indeed rise to operators local in $x$-space. Hence we
study first a class of operators local in field space. The result will be
that as a candidate for $X_0$ only $K_0$ survives whose study will be performed
in \cite{SiboldEden}. In the present ``basic case'' $X_0$ and subsequently also $X_j$ will be permitted to be non-local in field space.\\ 

\subsection{Local operators}

We know that $P_{\mu}$, $D$, $M_{\mu\nu}$ and $K_{\mu}$ are local operators
(s.\ appendix for explicit expressions).
But are there other local operators ? We will answer this question by 
considering rotationally invariant objects (as $P_0$, $D$ and $K_{0}$ are). Let's consider :
\begin{align}
 A(a,b,l_1,l_2,l_3,l_4,l_5)=&\int{}d^3x~ x_0^{l_5}(x_mx^m)^{l_1}(x^l)^{l_3}(x^m)^{l_4}\partial_0^a\partial_j^{l_2}\partial_l^{l_3}\phi(x)\partial_0^b(\partial^j)^{l_2}\partial_m^{l_4}\phi(x)
\end{align}
A local expression in terms of fields is a sum of such $A(a,b,l_1,l_2,l_3,l_4,l_5)$ (which is rotationally invariant). In fact these A's are not independent since some of them are related by partial integration but we will consider those since the expression is symmetric as far as the two fields are concerned. We consider only products of two fields since we want an operator with terms having one product of $a_{\mathbf{p}}^{\dagger}$ (or of its derivatives) and $a_{\mathbf{p}}$ (or of its derivatives). We will proceed as follows. We
consider a general linear combination of such monomials $A$
\begin{equation}
  \mathcal{A} =\sum_{a,b,l_1,l_2,l_3,l_4,l_5}a_{a,b,l_1,l_2,l_3,l_4,l_5}A(a,b,l_1,l_2,l_3,l_4,l_5)
\end{equation}
the $a_{a,b,l_1,l_2,l_3,l_4,l_5}$ being complex coefficients. Then we replace their expression in terms of fields by their expressions in terms of creation and annihilation operators. This gives us a sum of terms with product of $a_{\mathbf{p}}^{\dagger}a_{\mathbf{p}}^{\dagger}$, $a_{\mathbf{p}}^{\dagger}a_{\mathbf{p}}$, $a_{\mathbf{p}}a_{\mathbf{p}}^{\dagger}$, $a_{\mathbf{p}}a_{\mathbf{p}}$ (or of their derivatives). We then require that all terms with $a_{\mathbf{p}}^{\dagger}a_{\mathbf{p}}^{\dagger}$ or $a_{\mathbf{p}}a_{\mathbf{p}}$ vanish. This gives us conditions on the coefficients $a_{a,b,l_1,l_2,l_3,l_4,l_5}$. We then also require that the coefficients don't depend on $x_0$ and that the resulting operator doesn't explicitly depend on $x_0$ except through $a_{\mathbf{p}}^{\dagger}$ and $a_{\mathbf{p}}$, which gives us other conditions on them. Then the result is a local, rotationally invariant operator. Unfortunately the calculations involved are not easy and thus we will restrict the sum : we will consider in the sum only A's with, at most, two x's (which corresponds to two derivatives on $a_{\mathbf{p}}^{\dagger}$ and $a_{\mathbf{p}}$). That is to say that the sum is a sum on $a,~b,~l_1,~l_2,~l_3,~l_4,~l_5$ with $l_5+2l_1+l_3+l_4\le2$. This is still pretty general since, e.g. all the rotationally invariant operators of the conformal algebra are obtained this way.\\

Our hope is that by using this method, one discovers a suitable $Q_0$. Since one has $[P_0,Q_0]=i$, the dimension of $Q_0$ is $-1$ (because the dimension of $P_0$ is 1). This means that in our sum, $a+b+2l_2-2l_1-l_5=0$. By performing the steps described above, one finds that a rotationally invariant local operator obtained by this method is necessarily proportional to $K_0$. 
On the one hand, this result provides us with a new definition of the conformal operator $K_0$ as being the only local operator of dimension -1 (within the prescribed set). On the other hand, this shows that all other coordinate operators which one wants to build are to be found amongst nonlocal operators.

\vspace{0.5cm}
As an aside,
out of curiosity, one can look at sums of other dimensions. If, for example the sum contains terms of dimension 0 ($a+b+2l_2-2l_1-l_5=1$), one finds by applying the method described above that a rotationally invariant operator of dimension 0 is necessarily proportional to D. Once again, this provides us with a new definition of D. When proceeding with dimension 1 ($a+b+2l_2-2l_1-l_5=2$), one finds two possible operators, $P_0$, of course, and another one :
\begin{align}
 R_0=\int{}d^3p~(5\beta-3\alpha) p_0p^ja_{\mathbf{p}}^{\dagger}\frac{\partial a_{\mathbf{p}}}{\partial p^j}+\alpha p_0^3a_{\mathbf{p}}^{\dagger}\frac{\partial^2 a_{\mathbf{p}}}{\partial p^j \partial p_j}+\beta p_0 p^j p^l a_{\mathbf{p}}^{\dagger}\frac{\partial^2 a_{\mathbf{p}}}{\partial p^j \partial p^l}~(\alpha,~\beta\in\mathbb{C})
\end{align}
and if one requires that $R_0$ is also hermitian, then $\alpha$ and $\beta$ are real.

\subsection{Permitted nonlocal coordinate operators}

Giving up locality for the coordinate operators forces us to use a general approach. Let's first try to define a $Q_0$. For simplicity we assume it to be rotationally invariant. Moreover, in analogy to the conformal algebra, we will consider the following operator :
\begin{align}
 S_0(\alpha,\beta,\gamma,\delta)=&\int{}d^3p~\frac{\alpha}{p_0} a_{\mathbf{p}}^{\dagger}a_{\mathbf{p}}+\beta \frac{p^j}{p_0}a_{\mathbf{p}}^{\dagger}\frac{\partial a_{\mathbf{p}}}{\partial p^j}+\gamma p_0 a_{\mathbf{p}}^{\dagger}\frac{\partial^2 a_{\mathbf{p}}}{\partial p^l \partial p_l}+\delta a_{\mathbf{p}}^{\dagger}\frac{p^l p^j}{p_0}\frac{\partial^2 a_{\mathbf{p}}}{\partial p^l \partial p^j}
\\ &(\mathrm{with}~\alpha,~\beta,~\gamma,~\delta \in \mathbb{C}~\mathrm{free~parameters})\nonumber
\end{align}
This is the most general form of a rotationally invariant operator with at most second order derivatives and of dimension $-1$. We calculate $[P_0,S_0]$ and have conditions on $\alpha,~\beta,~\gamma,~\delta$ in order to satisfy the commutator (\ref{precooopfin}). One soon finds that having $[P_0,S_0]=i$ is not directly possible. But one can choose some particular coefficients $\alpha,~\beta,~\gamma,~\delta$ so that $[P_0,S_0]=iN$ with $N=\int{}d^3p~a_{\mathbf{p}}^{\dagger}a_{\mathbf{p}}$ the number operator
(every a n-particle state is an eigenvector of $N$ with eigenvalue n). Then one just defines $Q_0=N^{-1/2}S_0N^{-1/2}$. So, if one requires $[P_0,S_0]=iN$, one has :
\begin{align}
S_0(A_0,B_0)=&\int{}d^3p~\frac{A_0-i}{p_0} a_{\mathbf{p}}^{\dagger}a_{\mathbf{p}}+(2B_0-i) \frac{p^j}{p_0}a_{\mathbf{p}}^{\dagger}\frac{\partial a_{\mathbf{p}}}{\partial p^j}+B_0 p_0 a_{\mathbf{p}}^{\dagger}\frac{\partial^2 a_{\mathbf{p}}}{\partial p^l \partial p_l}+B_0 a_{\mathbf{p}}^{\dagger}\frac{p^l p^j}{p_0}\frac{\partial^2 a_{\mathbf{p}}}{\partial p^l \partial p^j}\label{possiblex0}
\end{align}
with $A_0~\mathrm{and}~B_0\in\mathbb{C}$. If one also wants $S_0$ to be hermitian, then $A_0~\mathrm{and}~B_0\in\mathbb{R}$. Since we want an operator as simple as possible, we take $B_0=0$ and define the pre-coordinate operator :
\begin{align}
 X_0=&\int{}d^3p~\frac{A_0-i}{p_0} a_{\mathbf{p}}^{\dagger}a_{\mathbf{p}}+(-i) \frac{p^j}{p_0}a_{\mathbf{p}}^{\dagger}\frac{\partial a_{\mathbf{p}}}{\partial p^j}~~~(A_0\in\mathbb{R})\label{x0}
\end{align}
and, finally
\begin{equation}
 Q_0=N^{-1/2}X_0N^{-1/2} = X_0N^{-1},
\end{equation}
since $X_0$ and $N$ commute. We define such a pre-coordinate operator and perform
calculations mainly in terms of $X_0$ instead of directly $Q_0$ since they are then simpler. To obtain the result with $Q_0$ instead of $X_0$ one will generally just have to divide by $N$, since $N$ commutes with most of the considered operators. The only explanation as to why one chooses $X_0$ to contain only first order derivatives is simplicity. Since restricting ourselves to this case will still permit us to obtain interesting results, it is not very harmful.\\
We now have to define in the same way the $X_j$ (on can see that the same thing as in the $X_0$ case happens namely that one will have to divide $X_j$ by $N$ in order to obtain a $Q_j$). In order to do this, we'll consider the following operator which is the most general operator of dimension $-1$, with at most second order derivatives and so that each term has one, and only one, index j so that $X_j$ can later be contracted properly :
\begin{eqnarray}
 S_j(\alpha_i)=\int{}d^3p~&\alpha_1\frac{1}{p^j}a_{\mathbf{p}}^{\dagger} a_{\mathbf{p}}+\alpha_2\frac{p_j}{p_0^2}a_{\mathbf{p}}^{\dagger} a_{\mathbf{p}}+\alpha_3 a_{\mathbf{p}}^{\dagger}\frac{\partial a_{\mathbf{p}}}{\partial p^j}+\alpha_4 \frac{p_j p^l}{p_0^2}a_{\mathbf{p}}^{\dagger} \frac{\partial a_{\mathbf{p}}}{\partial p^l}+\alpha_5\frac{p^l}{p^j}a_{\mathbf{p}}^{\dagger} \frac{\partial a_{\mathbf{p}}}{\partial p^l} \nonumber
\\&+\alpha_6 p_j a_{\mathbf{p}}^{\dagger}\frac{\partial^2 a_{\mathbf{p}}}{\partial p^l \partial p_l}
 +\alpha_7 p^l a_{\mathbf{p}}^{\dagger} \frac{\partial^2 a_{\mathbf{p}}}{\partial p^l \partial p^j}+\alpha_{8} \frac{p_j p^l p^m}{p_0^2} a_{\mathbf{p}}^{\dagger} \frac{\partial^2 a_{\mathbf{p}}}{\partial p^l \partial p^m}+\alpha_9\frac{p_0^2}{p^j}a_{\mathbf{p}}^{\dagger}\frac{\partial^2 a_{\mathbf{p}}}{\partial p^l \partial p_l}\nonumber
\\ &+\alpha_{10} \frac{p^l p^m}{p^j} a_{\mathbf{p}}^{\dagger} \frac{\partial^2 a_{\mathbf{p}}}{\partial p^l \partial p^m}
\\ (\alpha_i\in\mathbb{C})\nonumber
\end{eqnarray}

By requiring that $[P_j,S_j]=-iN$, there are some conditions on the $\alpha_i$ and one has now to consider :
\begin{align}
 S_j=\int{}d^3p~&\beta_1\frac{1}{p^j}a_{\mathbf{p}}^{\dagger} a_{\mathbf{p}}+\beta_2\frac{p_j}{p_0^2}a_{\mathbf{p}}^{\dagger} a_{\mathbf{p}}+\beta_3 a_{\mathbf{p}}^{\dagger}\frac{\partial a_{\mathbf{p}}}{\partial p^j}+(\beta_3+i)\frac{p^l}{p^j}a_{\mathbf{p}}^{\dagger} \frac{\partial a_{\mathbf{p}}}{\partial p^l}+\beta_4 \frac{p^l p^m}{p^j} a_{\mathbf{p}}^{\dagger} \frac{\partial^2 a_{\mathbf{p}}}{\partial p^l \partial p^m}\nonumber
\\ &~-2\beta_4 p^l a_{\mathbf{p}}^{\dagger} \frac{\partial^2 a_{\mathbf{p}}}{\partial p^l \partial p^j}+\beta_4 p_j a_{\mathbf{p}}^{\dagger}\frac{\partial^2 a_{\mathbf{p}}}{\partial p^l \partial p_l}\hbox{\hskip1cm} (\beta_i\in\mathbb{C})
\end{align}
But we also want $S_j$ to be hermitian which restricts further the coefficients. 
And so the most general form of an hermitian operator of dimension $-1$ with one, and only one, index j per term, with at most second order derivatives and which verifies the commutation relation $[P_j,S_j]=-iN$ is :
\begin{align}
 S_j=\int{}d^3p~\left(A_j+i(C_j+1)\right)\frac{1}{p^j}a_{\mathbf{p}}^{\dagger} a_{\mathbf{p}}+B_j\frac{p_j}{p_0^2}a_{\mathbf{p}}^{\dagger} a_{\mathbf{p}}+iC_ja_{\mathbf{p}}^{\dagger}\frac{\partial a_{\mathbf{p}}}{\partial p^j}+i(C_j+1)\frac{p^l}{p^j}a_{\mathbf{p}}^{\dagger} \frac{\partial a_{\mathbf{p}}}{\partial p^l}\label{possiblexj}
\end{align}
with $A_j,~B_j,~C_j\in\mathbb{R}$. Since there is no reason to favor one direction, the three coefficients $A_j,~B_j~\mathrm{and}~C_j$ don't depend on j. Once again, we want a simple pre-coordinate operator $X_j$ and one can see by calculating commutators with other operators that the terms with $\frac{1}{p^j}$ complicate those commutators. So one will consider $C_j=-1$ and $A_j=0$. Moreover, to simplify further this operator, we will require that all $X_j$ commute with each other which leads to $B_j=0$. This finally give us the $X_j$ that we will consider:
\begin{align}
 X_j=&\int{}d^3p~(-i)a_{\mathbf{p}}^{\dagger} \frac{\partial a_{\mathbf{p}}}{\partial p^j}\label{xj}
\end{align}
And one defines 
\begin{equation}
Q_j=N^{-1/2}X_jN^{-1/2}= X_jN^{-1}.
\end{equation}
One should note the fact that these operators are not defined on the vacuum since $N$ can't be inverted on it. Hence we have to have a look at the domains, where  the $Q_{\mu}$ are defined.

We will now give a way to formally express the $Q_{\mu}$ without having to use $N^{-1}$ or rather, we will give an explicit writing for $N^{-1}$. For that matter one introduces the ``cut'' operator $C_n$ :
\begin{eqnarray}
 C_n= \nonumber
\\&{\hskip-1cm}\left(\int{}d^3p_1...d^3p_n~\frac{1}{n!}a_{\mathbf{p_1}}^{\dagger}...a_{\mathbf{p_n}}^{\dagger}a_{\mathbf{p_n}}...a_{\mathbf{p_1}}\right)\left(\mathbf{1}-\left(\int{}d^3p_1...d^3p_{n+1}~\frac{1}{(n+1)!}a_{\mathbf{p_1}}^{\dagger}...a_{\mathbf{p_{n+1}}}^{\dagger}a_{\mathbf{p_{n+1}}}...a_{\mathbf{p_1}}\right)\right)
\nonumber\\&
\end{eqnarray}
It is so called because one has $C_n|\mathrm{m-particles~state}>=\delta_n^m|\mathrm{m-particles~state}>$. 
I.e.\ this is a projector on n-particles states. Then one can define :
\begin{align}
Q_{\mu}=&\sum_{n=1}^{+\infty}C_n\frac{X_{\mu}}{n}C_n
\end{align}
We put two $C_n$ in order to have a projector on kets and bras.
And one also has :
\begin{align}
 X_{\mu}=&\sum_{n=1}^{+\infty}C_n(nQ_{\mu})C_n
\end{align}

\section{Properties of $X_\mu$ and $Q_\mu$}

\subsection{Lorentz covariance of the pre-coordinate operators}

By our above choice of $X_\mu$ instead of $K_\mu$ we gave up locality in
field space. What about Lorentz covariance? \\
For example, one has :
\begin{align}
 [X_0,M_{j0}]=&\int{}d^3p~i\frac{p_j}{p_0}\left(\frac{A_0-i}{p_0}a_{\mathbf{p}}^{\dagger}a_{\mathbf{p}}-i\frac{p^l}{p_0}a_{\mathbf{p}}^{\dagger}\frac{\partial a_{\mathbf{p}}}{\partial p^l}\right)\neq -iX_j
\end{align}
Hence the $X_{\mu}$ do not form a fourvector, Lorentz covariance is broken.
Is this breaking due to our simplifications? Indeed, one can calculate $[S_0,M_{j0}]$ with $S_0$ given by (\ref{possiblex0}) that is to say the pre-coordinate operator before our simplifications. Then one has :
\begin{align}
 [S_0,M_{j0}]=i\int{}d^3p~&(B_0+A_0-i)\frac{p_j}{p_0^2}a_{\mathbf{p}}^{\dagger} a_{\mathbf{p}}+3B_0 a_{\mathbf{p}}^{\dagger}\frac{\partial a_{\mathbf{p}}}{\partial p^j}+(3B_0-i) \frac{p_j p^l}{p_0^2}a_{\mathbf{p}}^{\dagger} \frac{\partial a_{\mathbf{p}}}{\partial p^l}\nonumber\\ &-B_0 p_j a_{\mathbf{p}}^{\dagger}\frac{\partial^2 a_{\mathbf{p}}}{\partial p^l \partial p_l}
+2B_0 p^l a_{\mathbf{p}}^{\dagger} \frac{\partial^2 a_{\mathbf{p}}}{\partial p^l \partial p^j}+B_0 \frac{p_j p^l p^m}{p_0^2} a_{\mathbf{p}}^{\dagger} \frac{\partial^2 a_{\mathbf{p}}}{\partial p^l \partial p^m}
\end{align}
And when one compares with (\ref{possiblexj}) that is to say the pre-coordinate operator before our simplification, one can see that no choice of coefficients can provide us with the commutation relation $[S_0,M_{j0}]=-iS_j$. And so the Lorentz covariance of the $X_{\mu}$ is broken independently of our simplifications.

Nevertheless, some commutators of $X_{\mu}$ with the Lorentz operators are still those that one would expect :
\begin{align}
 [X_0,M_{jk}]=&0
\\ [X_j,M_{lm}]=&i\delta_{jl}X_m-i\delta_{jm}X_l
\end{align}
to be compared with :
\begin{align}
 [K_0,M_{jk}]=&0 & [P_0,M_{jk}]=&0\nonumber
\\ [K_j,M_{lm}]=&i\delta_{jl}K_m-i\delta_{jm}K_l & [P_j,M_{lm}]=&i\delta_{jl}P_m-i\delta_{jm}P_l\nonumber
\end{align}

We still haven't spoken about the commutator $[X_j,M_{l0}]$. One would expect it to be equal to $i\delta_{jl}X_0$ (since $[P_j,M_{l0}]=i\delta_{jl}P_0$ and $[K_j,M_{l0}]=i\delta_{jl}K_0$). In fact one has :
\begin{align}
 [X_j,M_{l0}]=&\int{}d^3p~\left(\delta_{jl}(\frac{1}{2p_0}+\frac{p_jp_l}{2p_0^3}\right)a_{\mathbf{p}}^{\dagger}a_{\mathbf{p}}+\frac{p_j}{p_0}a_{\mathbf{p}}^{\dagger}\frac{\partial a_{\mathbf{p}}}{\partial p^l}
\end{align}
which, at first sight, seems far from $i\delta_{jl}X_0$. But if one
contracts the indices, one has $[X^j,M_{j0}]=iX_0+A_0N_{-1}$ with
$N_{-1}=\int{}d^3p~\frac{1}{p_0}a_{\mathbf{p}}^{\dagger}a_{\mathbf{p}}$.
So if one takes $A_0=0$, one has $[X^j,M_{j0}]=iX_0$ to be compared
with $[P^j,M_{j0}]=3iP_0$ and $[K^j,M_{j0}]=3iK_0$. So even if one
doesn't have a true covariance, one can still obtain $X_0$ back starting
from the $X_j$ and using the Lorentz operators. Unfortunately, as we have
seen, the contrary is not true, one can't obtain the $X_j$ starting from
$X_0$ and using the Lorentz operators. Even if these calculations hint
us to take $A_0=0$ (and we will find below an other good reason for doing
so), for the time being we will continue to use it as a parameter since
the Lorentz covariance is broken anyway and this parameter might give
us some additional freedom.\\
Before proceeding let us give a comment on this result that Lorentz
covariance is necessarily broken at the level of the (pre-)coordinate
operators. In the next
section we shall see that time and space operators do not commute, as
we indeed are looking for. We may thus compare this situation with
other approaches to a non-commutative spacetime like the construction via
Moyal products \cite{DoplicherFredenhagenRoberts}. There too, Lorentz
covariance is lost. As will be clear from considerations in sect.\ 5
the loss of Lorentz covariance in our case is however not necessarily the
last word on this subject because it happens on the level of
(pre-)coordinate operators and covariance might be reestablished via suitably
constructed functions of them.\\

\subsection{Algebra of the pre-coordinate operators}

Now that we have finally obtained coordinate operators (or rather pre-coordinate operators) that verify the commutation relations (\ref{precooopfin}), we have to understand them. The first thing one should calculate is the commutation relation between them. And so one has :
\begin{align}
 [X_0,X_j]=&\int{}d^3p~i(A_0-i)\frac{p_j}{p_0^3}a_{\mathbf{p}}^{\dagger}a_{\mathbf{p}}+\frac{1}{p_0}a_{\mathbf{p}}^{\dagger}\frac{\partial a_{\mathbf{p}}}{\partial p^j}+\frac{p_jp^l}{p_0^3}a_{\mathbf{p}}^{\dagger}\frac{\partial a_{\mathbf{p}}}{\partial p^l}
\\ [X_j,X_l]=&0
\end{align}
We required the $X_j$ to commute with each other by an argument of simplicity. But as one can see, $X_0$ and $X_j$ don't commute (and can't be made to commute by taking different coefficients in (\ref{possiblex0}) and in (\ref{possiblexj})). This means that there won't be common eigenstates for the four $X_{\mu}$. We are satisfied with this non-commutativity since if they were commuting, their eigenvalues would certainly have a relation between them as there is for the energy-momentum operators or the conformal operators. And since we want to give the meaning of coordinates to these eigenvalues, the time coordinate should not be directly linked to the spatial coordinate.
Incidentally we remark that for $A_0=0$ the operators $X_\mu$ coincide with those which
have been called $X_\mu(N)$ in \cite{Sibold}. There the commutators $[X_\mu,X_\nu]$ were
stated to be vanishing; this is wrong: the statement was due to a
calculational error.\\
Before solving the eigenvalue problems for these operators, let's look at the commutation relations of $X_{\mu}$ with the conformal algebra.

\subsection{Conformal covariance of the pre-coordinate operators}

\noindent
Let's first calculate the commutator with the dilatation operator :
\begin{align}
[X_{\mu},D]=-iX_{\mu}
\end{align}
This is what one would have expected. Indeed, in the conformal algebra, for an operator $O$, one has $[O,D]=idO$, d being the dimension of the operator $O$. Here, $X_{\mu}$ being of dimension $-1$, the commutator is of this form. One can say that $X_{\mu}$ transforms properly under dilatations.

Let's now look at the commutators with the energy-momentum operators. We already know that $[P_{\mu},X_{\mu}]=i\eta_{\mu\mu}N$ since this was the requirement to define the $X_{\mu}$. The others are :
\begin{align}
 [P_j,X_l]=&0\qquad (j\neq l)
\\ [P_j,X_0]=&\int{}d^3p~i\frac{p_j}{p_0}a_{\mathbf{p}}^{\dagger}a_{\mathbf{p}}\equiv i\frac{P_j}{P_0}\neq iP_jP_0^{-1}
\\ [P_0,X_j]=&\int{}d^3p~i\frac{p_j}{p_0}a_{\mathbf{p}}^{\dagger}a_{\mathbf{p}}\equiv i\frac{P_j}{P_0}\neq iP_jP_0^{-1}
\end{align}
$\frac{P_j}{P_0}$ being only a notation and being the same as $P_jP_0^{-1}$ only on one particle-states. As one can see, the spatial energy-momentum operators commute with the spatial pre-coordinate operators with different index. One can interpret this as the fact that the motion in one direction doesn't interfere with the coordinates of other directions. The non-commutativity of $X_0$ with $P_j$ and that of $P_0$ with $X_j$ will later give rise to some technical inconvenience when trying to define fields.

Finally we look at the commutators between $X_{\mu}$ and the special conformal operators. One has :
\begin{align}
 [X_0,K_0]=&\int{}d^3p(\frac{i}{4}-A_0)\frac{1}{p_0^2}a_{\mathbf{p}}^{\dagger}a_{\mathbf{p}}+(4i-2A_0)\frac{p^l}{p_0^2}a_{\mathbf{p}}^{\dagger}\frac{\partial a_{\mathbf{p}}}{\partial p^l}+ia_{\mathbf{p}}^{\dagger}\frac{\partial^2 a_{\mathbf{p}}}{\partial p^l \partial p_l}+2i\frac{p^j p^l}{p_0^2}a_{\mathbf{p}}^{\dagger}\frac{\partial^2 a_{\mathbf{p}}}{\partial p^j \partial p^l}
\\ [X_0,K_j]=&\int{}d^3p~(\frac{3i}{4}-A_0)\frac{p_j}{p_0^3}a_{\mathbf{p}}^{\dagger}a_{\mathbf{p}}+(i-2A_0)\frac{1}{p_0}a_{\mathbf{p}}^{\dagger}\frac{\partial a_{\mathbf{p}}}{\partial p^j}+i\frac{p_jp^l}{p_0^3}a_{\mathbf{p}}^{\dagger}\frac{\partial a_{\mathbf{p}}}{\partial p^l}+i\frac{p_j}{p_0}a_{\mathbf{p}}^{\dagger}\frac{\partial^2 a_{\mathbf{p}}}{\partial p^l \partial p_l}
\\ [X_j,K_0]=&\int{}d^3p~\frac{3i}{4}\frac{p_j}{p_0^3}a_{\mathbf{p}}^{\dagger}a_{\mathbf{p}}+i\frac{p_jp^l}{p_0^3}a_{\mathbf{p}}^{\dagger}\frac{\partial a_{\mathbf{p}}}{\partial p^l}+i\frac{p_j}{p_0}a_{\mathbf{p}}^{\dagger}\frac{\partial^2 a_{\mathbf{p}}}{\partial p^l \partial p_l}
\\ [X_j,K_l]=&\int{}d^3p~\left(\frac{i}{4}(\frac{\delta_{jl}}{p_0^2}+\frac{2p_jp_l}{p_0^4})\right)a_{\mathbf{p}}^{\dagger}a_{\mathbf{p}}+2ia_{\mathbf{p}}^{\dagger}\frac{\partial^2 a_{\mathbf{p}}}{\partial p^j \partial p^l}-i\delta_{jl}a_{\mathbf{p}}^{\dagger}\frac{\partial^2 a_{\mathbf{p}}}{\partial p^m \partial p_m}
\end{align}
These commutators have no direct interpretation since the result of a commutator of $X_{\mu}$ with $K_{\nu}$ is an operator of dimension $-2$ which is then out of the conformal algebra.

\section{Eigenvalue problems and reconstruction of Fock space}

\subsection{Eigenvalue problem of $X_\mu$ and $Q_\mu$}

Let's now look at the eigenvalue problem for $X_{\mu}$. As we have already seen, the $X_{\mu}$ are non-commuting and this will lead us to \textit{two} different eigenvalue problems. We will start by first looking at the $X_{0}$ eigenvalue problem. We consider a state $|q_0>=\int{}d^3p~f(\vec{p},q_0)|\vec{p}>$ and solve a partial differential equation on f so that $|q_0>$ is an eigenstate of $X_{0}$. The partial differential equation obtained is :
\begin{align}
 \frac{A_0-i}{p_0}f-i\frac{p^l}{p_0}\frac{\partial f}{\partial p^l}=q_0f
\end{align}
with $q_0$ the eigenvalue. We can use the following ansatz : $f(\vec{p},q_0)=g(p_0,q_0)Y(\theta,\phi,q_0)$ and the result doesn't depend on $Y(\theta,\phi,q_0)$ ($\theta$ and $\phi$ being, as above, the angle of $\vec{p}$ in spherical coordinates). This ansatz gives us an ordinary first order differential equation on $g$. Once one has solved it, one has the following eigenstates for $X_0$ :
\begin{align}\label{q0eigenst}
 |q_0>=&\int{}d^3p~p_0^{-i(A_0-i)}e^{ip_0q_0}Y(\theta,\phi,q_0)|\vec{p}>
\\ X_0|q_0>=&q_0|q_0>\nonumber
\end{align}
with $Y(\theta,\phi,q_0)$ an arbitrary function. In view of the interpretation
of $Q_0$ as time operator we chose $q_0$ to be real. This is an additional assumption since $X_0$ is unbounded.\\
We will next look at the eigenvalue problem for $X_j$. Since they are commuting with each other, they have common eigenstates. The eigenvalue problem is the following :
\begin{align}
 \forall j,~~~~-i\frac{\partial f}{\partial p^j}=q_jf
\end{align}
And this give us the following eigenstates :
\begin{align}\label{qj-eigenst}
 |\vec{q}>=&\int{}d^3p~\frac{e^{i\vec{q}.\vec{p}}}{(2\pi)^{3/2}}|\vec{p}>
\\ X_j|\vec{q}>=&q_j|\vec{q}>\nonumber
\end{align}

the $\frac{1}{(2\pi)^{3/2}}$ being included to be coherent with the Fourier transform definition. We assume $q_j$ also to be real.\\
Once one knows the eigenstates of the $X_{\mu}$ in an $n$-particle subspace
one
can easily calculate the eigenstates of $Q_{\mu}$ in the same subspace.\\
What about completeness and normalization of the $Q_\mu\,$-eigenstates? For those of $Q_j$ an immediate answer is given by the simple result (\ref{qj-eigenst}):
we just invert the basis and express the $|\vec{p}>$
states in terms of the $|\vec{q}>$ states :
\begin{align}
 |\vec{p}>=&\int{}d^3q~\frac{e^{-i\vec{p}.\vec{q}}}{(2\pi)^{3/2}}|\vec{q}>
\end{align}
In this respect, $|\vec{p}>$ and $|\vec{q}>$ are the Fourier transform of each other and thus the $|\vec{q}>$ enjoy the same completeness and normalization properties as the $|\vec{p}>$. This holds also true for $n$-particle states.\\
The analogous considerations for the $Q_0$-eigenstates require a more detailed
treatment. Calculating  $<q'_0|q_0>$ from (\ref{q0eigenst}) one finds a $\delta_{+}$-function of the argument $q'_0-q_0$ and not a $\delta$-function. The difference, being
an imaginary contribution (principal value), suggests to consider the complex conjugate eigenfunction $g^{*}(p_0,q_0)$ for which one finds
\begin{equation}
g^{*}(p_0,q_0) = g(p_0,-q_0),
\end{equation}
if $A_0=0$ ! Hence ``time reversal'' eigenstates could perhaps show the
desired continuum normalization. Indeed, defining
\begin{equation}
|q_0>_{\pm} =\frac{1}{\sqrt{2}}(|q_0> \pm |-q_0>),
\end{equation}
first, the states $|q_0>_{\pm}$ are even, resp.\ odd under time reversal, defined by
$q_0 \rightarrow -q_0$ and second, they have the correct continuum
normalization
\begin{equation}
_{\pm}<q'_0,\theta_{q'},\phi_{q'}|q_0, \theta_q,\phi_q>_{\pm} =
      \frac{1}{q_0^2}\delta(q'_0-q_0)\delta(\phi_{q'}-\phi_q)\delta(cos\theta_{q'}-
                                               cos\theta_q).
\end{equation}  
Here we anticipated that $q_0$ can be restricted to $0\le q_0 \le\infty$ and labelled the degeneracy of the eigenstates belonging to $q_0$ suitably, in terms of angles.
The time reversal even/odd eigenstates read then explicitly
\begin{equation}
|q_0,\theta_q, \phi_q>_{\pm} = \int_0^\infty dp_0 p_0^2\int_0^{2\pi} d\phi_p \int d(cos\theta_p) g_{\pm}(p_0,q_0) Y(\theta_p,\phi_p;\theta_,\phi_q) |p_0,\theta_p,\phi_p>,
\end{equation}
where
\begin{eqnarray}
g_{\pm}(p_0,q_0)& = &\frac{1}{\sqrt{2\pi}} 
               \frac{e^{ip_0q_0} \pm e^{-ip_0q_0}}{p_0q_0}\\
Y(\theta_p,\phi_p;\theta_q,\phi_q) & = &\delta(\phi_p-\phi_q)
                                 \delta(cos\theta_p-cos\theta_q)\\
\end{eqnarray}
Inversely, the momentum eigenstates can be expressed, respectively in terms of either
$|q_0,...>_{+}$ or $|q_0,...>_{-}$ as
\begin{equation}
|p_0,\theta_p, \phi_p> = \int_0^\infty dq_0 q_0^2\int_0^{2\pi} d\phi_q \int_0^\pi d(cos\theta_q) g^*_{\pm}(p_0,q_0) Y(\theta_p,\phi_p;\theta_,\phi_q) |q_0,\theta_q,\phi_q>.
\end{equation}
These equations show that the one-particle states $|q_0,...>$ are
complete. By constructing
tensorproducts this holds then also for all $n$-particle states
($n\ne 0$).\\
The above analysis of eigenstates of $Q_0$ and the necessity to go
over to time reversal eigenstates which are thus no longer eigenstates
of $Q_0$ signal a possible way how to circumvent Paulis theorem (\cite{Pauli}): if
only these states are legitimate states to be used in all applications
of $Q_0$ then from its spectrum $q_0\in\mathbb{R}$ only the non-negative
portion is actually active. In the context of concrete examples it has
to be checked that this general conjecture indeed materializes.\\ 
 
\subsection{Basis independence}
The basic commutation rules for the operators $X_\mu$ and consequently those for
the $Q_\mu$ have been calculated from the explicit expressions in terms of
creation and annihilation operators. Hence they should hold on every state of Fock space (except the vacuum). However, $X_\mu$ and $Q_\mu$ are unbounded and obviously Hermitian, but not necessarily self-adjoint since their
eigenstates have only continuum normalization and we have not yet studied
their domains of definition. In order to go a step into this latter
direction we checked explicitly that at least the commutators (\ref{precooopfin})
hold indeed true on the bases $|\vec{p}>$, $|q_j>$, $|q_0>_{\pm}$.

\subsection{Fock space in terms of  pre-coordinate operator eigenstates}

We will now show that the Fock space which has been expressed until now in terms of the energy-momentum operator eigenstates can be rewritten with the pre-coordinate operator eigenstates. One defines a translation operator in $q$-space :
\begin{align}
 T_{\mathbf{q}}=&\int{}d^3p~e^{i\vec{p}.\vec{q}}a_{\mathbf{p}}^{\dagger}a_{\mathbf{p}} & T_{\mathbf{q}}|\vec{q'}>=&|\vec{q}+\vec{q'}>
\end{align}
We can then provide the one-particle-subspace of Fock space with bases which are conjugate to each other, $|\vec{p}>$ and $|\vec{q}>$ by defining :
\begin{align}
 a_{\mathbf{q}}^{\dagger}=&\int{}d^3p~\frac{e^{i\vec{p}.\vec{q}}}{(2\pi)^{3/2}}a_{\mathbf{p}}^{\dagger} & a_{\mathbf{q}}^{\dagger}|0>=&|\vec{q}>
\\ a_{\mathbf{q}}=&\int{}d^3p~\frac{e^{-i\vec{p}.\vec{q}}}{(2\pi)^{3/2}}a_{\mathbf{p}} & a_{\mathbf{q}}|\vec{q'}>=&\delta(\vec{q}-\vec{q'})
\\ T_{\mathbf{p}}=&\int{}d^3q~e^{-i\vec{p}.\vec{q}}a_{\mathbf{q}}^{\dagger}a_{\mathbf{q}} & T_{\mathbf{p}}|\vec{p'}>=&|\vec{p}+\vec{p'}>
\\ [a_{\mathbf{q}},a_{\mathbf{q'}}^{\dagger}]=&\delta(\vec{q}-\vec{q'}) & &
\end{align}

Having chosen the simplest expression for $X_j$, we are led to a really simple rewriting of the Fock space in terms of eigenstates of the spatial pre-coordinate operators. This is more or less a Fourier transform of the one-particle Fock space parametrized by the eigenstates of the energy-momentum operators. The extension to
the $n$-particle space ($n\ne0$) is obvious.

\vspace{1cm}

\section{Application: S-matrices and fields}

\subsection{Shift operator}

Before building S-matrices and fields, we will derive some useful relations by defining a shift operator that can be seen as a generalization of the above translation operator $T_{\mathbf{p}}$. Indeed we will consider the following operator : $e^{i\vec{X}.\vec{p}}$ ($\vec{p}$ being an arbitrary parameter) and show that it has some nice properties that will help us in constructing more complicated objects. For example one has :
\begin{align}
 e^{i\vec{X}.\vec{p'}}|\vec{p_1},...,\vec{p_n}>=&|\vec{p_1}-\vec{p'},...,\vec{p_n}-\vec{p'}>
\end{align}
In this sense it is truly a shift operator. It shifts all momenta by a fixed amount whereas if one uses the operator $T_{\mathbf{-p}}$ on a n-particles state, one has :
\begin{align}
 T_{\mathbf{-p'}}|\vec{p_1},...,\vec{p_n}>=|\vec{p_1}-\vec{p'},...,\vec{p_n}>+|\vec{p_1},\vec{p_2}-\vec{p'},...,\vec{p_n}>+...+|\vec{p_1},...,\vec{p_n}-\vec{p'}>\nonumber
\end{align}
$T_{\mathbf{p}}$ shifts the momentum of one particle whereas  $e^{i\vec{X}.\vec{p}}$ shifts the whole momentum space. Moreover, $e^{i\vec{X}.\vec{p}}$ is unitary.

We will now derive another useful expression involving $e^{i\vec{X}.\vec{p}}$. First of all one has :
\begin{align}
 [A,B^n]=&\sum_{k=0}^{n-1}\sum_{l=0}^{n-1-k}\binom{n-1-k}{l}B^{k+l}[A,B]^{(n-k-l)}\nonumber\\
          &\qquad(\mathrm{with}~[A,B]^{(n)}=[[[A,B],B,...,B])\nonumber
\\ \Rightarrow [\int{}d^3p~f(\vec{p})a_{\mathbf{p}}^{\dagger}a_{\mathbf{p}},X_j^n]=&\sum_{k=0}^{n-1}\sum_{l=0}^{n-1-k}\binom{n-1-k}{l}X_j^{k+l}i^{n-k-l}\int{}d^3p~\frac{\partial^{n-k-l} f(\vec{p})}{{\partial p^j}^{n-k-l}}a_{\mathbf{p}}^{\dagger}a_{\mathbf{p}}\nonumber
\end{align}

Then one uses this result by expressing the exponential in a power series :
\begin{align}
 [\int{}d^3p~f(\vec{p})a_{\mathbf{p}}^{\dagger}a_{\mathbf{p}},e^{iX_jp'_j}]=&[\int{}d^3p~f(\vec{p})a_{\mathbf{p}}^{\dagger}a_{\mathbf{p}},\sum_{n=0}^{+\infty}\frac{i^n}{n!}X_j^n{p'_j}^n]\nonumber
\\ =&\sum_{n=0}^{+\infty}\sum_{k=0}^{n-1}\sum_{l=0}^{n-1-k}\frac{i^n{p'_j}^n}{n!}\binom{n-1-k}{l}X_j^{k+l}i^{n-k-l}\int{}d^3p~\frac{\partial^{n-k-l} f(\vec{p})}{{\partial p^j}^{n-k-l}}a_{\mathbf{p}}^{\dagger}a_{\mathbf{p}}\nonumber
\\(k+l\rightarrow m)~~~ =&\sum_{n=0}^{+\infty}\sum_{m=0}^{n-1}\frac{i^n{p'_j}^n}{n!}i^{n-m}X_j^{m}\int{}d^3p~\frac{\partial^{n-m} f(\vec{p})}{{\partial p^j}^{n-m}}a_{\mathbf{p}}^{\dagger}a_{\mathbf{p}}\left(\sum_{r=0}^{m}\binom{n-1-r}{m-r}\right)\nonumber
\\ =&\sum_{n=0}^{+\infty}\sum_{m=0}^{n-1}\frac{i^n{p'_j}^n}{n!}i^{n-m}X_j^m\int{}d^3p~\frac{\partial^{n-m} f(\vec{p})}{{\partial p^j}^{n-m}}a_{\mathbf{p}}^{\dagger}a_{\mathbf{p}}\left(\frac{n}{n-m}\binom{n-1}{m}\right)\nonumber
\\=&\sum_{n=0}^{+\infty}\sum_{m=0}^{n-1}\frac{i^m{p'_j}^m}{m!}X_j^m\frac{(-p'_j)^{n-m}}{(n-m)!}\int{}d^3p~\frac{\partial^{n-m} f(\vec{p})}{{\partial p^j}^{n-m}}a_{\mathbf{p}}^{\dagger}a_{\mathbf{p}}\nonumber
\\=&\sum_{m=0}^{+\infty}\sum_{n=m+1}^{+\infty}\frac{i^m{p'_j}^m}{m!}X_j^m\frac{{(-p'_j)}^{n-m}}{(n-m)!}\int{}d^3p~\frac{\partial^{n-m} f(\vec{p})}{{\partial p^j}^{n-m}}a_{\mathbf{p}}^{\dagger}a_{\mathbf{p}}\nonumber
\\(n-m\rightarrow n)~~~=&\left(\sum_{m=0}^{+\infty}\frac{i^m{p'_j}^mX_j^m}{m!}\right)\left(\sum_{n=1}^{+\infty}\frac{{p'_j}^n}{n!}\int{}d^3p~\frac{\partial^n f(\vec{p})}{\partial {p_j}^n}a_{\mathbf{p}}^{\dagger}a_{\mathbf{p}}\right)\nonumber
\\ =&e^{iX_jp'_j}\int{}d^3p~(f(\vec{p}+p'_j\vec{e_j})-f(\vec{p}))a_{\mathbf{p}}^{\dagger}a_{\mathbf{p}}\nonumber
\end{align}
And so one has :
\begin{align}
 [\int{}d^3p~f(\vec{p})a_{\mathbf{p}}^{\dagger}a_{\mathbf{p}},e^{i\vec{X}.\vec{p'}}]=&e^{i\vec{X}.\vec{p'}}\int{}d^3p~(f(\vec{p}+\vec{p'})-f(\vec{p}))a_{\mathbf{p}}^{\dagger}a_{\mathbf{p}}
\end{align}
One can remark that the above expression simplifies greatly if $f$ is
linear (e.g. relevant, if one calculates the commutator of $P_j$
with $e^{i\vec{X}.\vec{p'}}$). Indeed, for a linear $f$ one has :
\begin{align}
 [\int{}d^3p~f(\vec{p})a_{\mathbf{p}}^{\dagger}a_{\mathbf{p}},e^{i\vec{X}.\vec{p'}}]=&e^{i\vec{X}.\vec{p'}}Nf(\vec{p'})~~~~~(f~\mathrm{linear})
\end{align}
And if one replaces $X_j$ by $Q_j$, the above relations become :
\begin{align}
 [\int{}d^3p~f(\vec{p})a_{\mathbf{p}}^{\dagger}a_{\mathbf{p}},e^{i\vec{Q}.\vec{p'}}]=&e^{i\vec{Q}.\vec{p'}}\int{}d^3p~(f(\vec{p}+\vec{p'}N^{-1})-f(\vec{p}))a_{\mathbf{p}}^{\dagger}a_{\mathbf{p}}\nonumber
\\ [\int{}d^3p~f(\vec{p})a_{\mathbf{p}}^{\dagger}a_{\mathbf{p}},e^{i\vec{Q}.\vec{p'}}]=&e^{i\vec{Q}.\vec{p'}}f(\vec{p'})~~~~~(f~\mathrm{linear})\nonumber
\end{align}
with $f(\vec{p}+\vec{p'}N^{-1})$ being $f(\vec{p}+\frac{\vec{p'}}{n})$ on a n-particles state. Once again, we can interpret $e^{i\vec{Q}.\vec{p}}$ as a shift operator. The commutation relations of $e^{i\vec{Q}.\vec{p}}$ with $P_j$ are then really simple :
\begin{align}
 [P_j,e^{i\vec{Q}.\vec{p}}]=&e^{i\vec{Q}.\vec{p}}p_j
\end{align}

Unfortunately, the commutation with $P_0$ is not as simple due to the fact that the dispersion law $p_0^2+p_jp^j=0$ is non-linear :
\begin{align}
 [P_0,e^{i\vec{Q}.\vec{p'}}]=&e^{i\vec{Q}.\vec{p'}}\int{}d^3p~(\omega_{(\mathbf{p}+N^{-1}\mathbf{p'})}-\omega_\mathbf{p})a_{\mathbf{p}}^{\dagger}a_{\mathbf{p}}
\end{align}

Similarly, one shows that :
\begin{align}
 [\int{}d^3p~f(\vec{p})a_{\mathbf{p}}^{\dagger}a_{\mathbf{p}},e^{i\vec{Q}.\vec{p'}}]=&\left(\int{}d^3p~(f(\vec{p})-f(\vec{p}-\vec{p'}N^{-1}))a_{\mathbf{p}}^{\dagger}a_{\mathbf{p}}\right)e^{i\vec{Q}.\vec{p'}}
\\ [\int{}d^3p~f(\vec{p})a_{\mathbf{p}}^{\dagger}a_{\mathbf{p}},e^{i{Q_0}{p'_0}}]=&e^{iQ_0{p'_0}}\int{}d^3p~(f(\vec{p}-\frac{p'_0}{p_0}\vec{p}N^{-1})-f(\vec{p}))a_{\mathbf{p}}^{\dagger}a_{\mathbf{p}}
 \\ [\int{}d^3p~f(\vec{p})a_{\mathbf{p}}^{\dagger}a_{\mathbf{p}},e^{iQ_0{p'_0}}]=&\left(\int{}d^3p~(f(\vec{p})-f(\vec{p}+\frac{p'_0}{p_0}\vec{p}N^{-1}))a_{\mathbf{p}}^{\dagger}a_{\mathbf{p}}\right)e^{iQ_0{p'_0}}
\end{align}

\subsection{Construction of S-matrices}

We will formally build S-matrices using the coordinate operators $Q_j$.
The shift operator will be the central tool to do so. We will then try
to give an interpretation of such an S-matrix as an S-matrix of a free
theory in curved space. We will in fact build the S-matrix by giving
the matrix elements between m-particles states and n-particle states.
We will give an example on the matrix elements of two-particles states
going into two-particles states, the generalization to all the other
matrix elements being easy. Let's consider :
\begin{align}
 S_{2,2}=&\frac{C_2}{2!}\left(\int{}d^3p_1d^3p_2{}~e^{i\vec{Q}.\vec{f}(\vec{p_1},\vec{p_2})}a_{\mathbf{p_1}}^{\dagger}e^{i\vec{Q}.\vec{g}(\vec{p_1},\vec{p_2})}a_{\mathbf{p_2}}^{\dagger}a_{\mathbf{p_1}}a_{\mathbf{p_2}}\right)\frac{C_2}{2!}
\\ =&\frac{C_2}{2!}\left(\int{}d^3p_1d^3p_2{}~e^{i\frac{\vec{X}.\vec{f}(\vec{p_1},\vec{p_2})}{2}}a_{\mathbf{p_1}}^{\dagger}e^{i\vec{X}.\vec{g}(\vec{p_1},\vec{p_2})}a_{\mathbf{p_2}}^{\dagger}a_{\mathbf{p_1}}a_{\mathbf{p_2}}\right)\frac{C_2}{2!}\nonumber
\end{align}
with $C_2$ the cut operator defined above. One puts it here to be sure that the only non-zero matrix elements for $S_{2,2}$ are really two-particles states going into two-particles states. One could get along with only one of them but there are two in order to have a two particle projector with respect to in- and out-states. Then one has :

\begin{align}
 <\vec{p'_3}\vec{p'_4}|S_{2,2}|\vec{p'_1}\vec{p'_2}>=&\frac{1}{(2!)(2!)}<\vec{p'_3}\vec{p'_4}|e^{i\frac{\vec{X}.\vec{f}(\vec{p'_1},\vec{p'_2})}{2}}a_{\mathbf{p'_1}}^{\dagger}e^{i\vec{X}.\vec{g}(\vec{p'_1},\vec{p'_2})}a_{\mathbf{p'_2}}^{\dagger}|0>\nonumber
\\&+\frac{1}{(2!)(2!)}<\vec{p'_3}\vec{p'_4}|e^{i\frac{\vec{X}.\vec{f}(\vec{p'_2},\vec{p'_1})}{2}}a_{\mathbf{p'_2}}^{\dagger}e^{i\vec{X}.\vec{g}(\vec{p'_2},\vec{p'_1})}a_{\mathbf{p'_1}}^{\dagger}|0>\nonumber
\\ =&\frac{1}{(2!)(2!)}<\vec{p'_3}\vec{p'_4}|e^{i\frac{\vec{X}.\vec{f}(\vec{p'_1},\vec{p'_2})}{2}}a_{\mathbf{p'_1}}^{\dagger}|\vec{p'_2}-\vec{g}(p'_1,p'_2)>\nonumber
\\&+\frac{1}{(2!)(2!)}<\vec{p'_3}\vec{p'_4}|e^{i\frac{\vec{X}.\vec{f}(\vec{p'_2},\vec{p'_1})}{2}}a_{\mathbf{p'_2}}^{\dagger}|\vec{p'_1}-\vec{g}(p'_2,p'_1)>\nonumber
\end{align}
\begin{align}
<\vec{p'_3}\vec{p'_4}|S_{2,2}|\vec{p'_1}\vec{p'_2}>=&\frac{1}{(2!)(2!)}\delta\left(\vec{p'_3}-\vec{p'_1}+\frac{\vec{f}(\vec{p'_1},\vec{p'_2})}{2}\right)\delta\left(\vec{p'_4}-\vec{p'_2}+\frac{\vec{f}(\vec{p'_1},\vec{p'_2})}{2}+\vec{g}(\vec{p'_1},\vec{p'_2})\right)\nonumber
\\&+\frac{1}{(2!)(2!)}\delta\left(\vec{p'_3}-\vec{p'_2}+\frac{\vec{f}(\vec{p'_1},\vec{p'_2})}{2}+\vec{g}(\vec{p'_1},\vec{p'_2})\right)\delta\left(\vec{p'_4}-\vec{p'_1}+\frac{\vec{f}(\vec{p'_1},\vec{p'_2})}{2}\right)\nonumber
\\&+\frac{1}{(2!)(2!)}\delta\left(\vec{p'_3}-\vec{p'_2}+\frac{\vec{f}(\vec{p'_2},\vec{p'_1})}{2}\right)\delta\left(\vec{p'_4}-\vec{p'_1}+\frac{\vec{f}(\vec{p'_2},\vec{p'_1})}{2}+\vec{g}(\vec{p'_2},\vec{p'_1})\right)\nonumber
\\&+\frac{1}{(2!)(2!)}\delta\left(\vec{p'_3}-\vec{p'_1}+\frac{\vec{f}(\vec{p'_2},\vec{p'_1})}{2}+\vec{g}(\vec{p'_2},\vec{p'_1})\right)\delta\left(\vec{p'_4}-\vec{p'_2}+\frac{\vec{f}(\vec{p'_2},\vec{p'_1})}{2}\right)\nonumber
\end{align}
And by defining $\vec{g'}(\vec{p_1},\vec{p_2})=-\vec{g}(\vec{p_1},\vec{p_2})-\frac{\vec{f}(\vec{p_1},\vec{p_2})}{2}+\vec{p_2}$ and $\vec{f'}(\vec{p_1},\vec{p_2})=-2\vec{f}(\vec{p_1},\vec{p_2})+2\vec{p_1}$, one has :
\begin{align}
<\vec{p'_3}\vec{p'_4}|S_{2,2}|\vec{p'_1}\vec{p'_2}>=&\frac{1}{4}\left(\delta(\vec{p'_3}-\vec{f'}(\vec{p'_1},\vec{p'_2}))\delta(\vec{p'_4}-\vec{g'}(\vec{p'_1},\vec{p'_2}))+\delta(\vec{p'_3}-\vec{g'}(\vec{p'_1},\vec{p'_2}))\delta(\vec{p'_4}-\vec{f'}(\vec{p'_1},\vec{p'_2}))\right.\nonumber
\\ &~~~~+\left.\delta(\vec{p'_3}-\vec{f'}(\vec{p'_2},\vec{p'_1}))\delta(\vec{p'_4}-\vec{g'}(\vec{p'_2},\vec{p'_1}))+\delta(\vec{p'_3}-\vec{g'}(\vec{p'_2},\vec{p'_1}))\delta(\vec{p'_4}-\vec{f'}(\vec{p'_2},\vec{p'_1}))\right)\nonumber
\end{align}
And so one has :
\begin{align}
 \begin{array}{ccc} & S_{2,2} & \\ |\vec{p'_1},\vec{p'_2}> & \longrightarrow & \frac{1}{2}\left(|\vec{f'}(\vec{p'_1},\vec{p'_2}),\vec{g'}(\vec{p'_1},\vec{p'_2})>+|\vec{f'}(\vec{p'_2},\vec{p'_1}),\vec{g'}(\vec{p'_2},\vec{p'_1})>\right)\end{array}
\end{align}
We obtain a superposition of two states at the end because the result has to be symmetric in $\vec{p'_1}$ and $\vec{p'_2}$. Since one can choose whatever functions $f'$ and $g'$ one wants, one is able to construct an arbitrary S-matrix. The generalization to an arbitrary matrix element (not just two-particles states going into two-particles states) is easy. This construction is entirely based on the properties of the shift operator. Since the shift operator is unitary, so is the S-matrix one constructs by this method.

\subsection{S-matrix in conformally flat and asymptotically flat space-time}

We now want to give to the S-matrix the meaning of an evolution operator of
states in curved space-time. In the algebraic approach to quantization
where one starts from an algebra of observables the
Gelfand-Naimark-Segal construction provides one in a natural way with
Fock spaces (\cite{Waldbook}, \cite{DimockKay}) and the principle of
general covariance
\cite{BrunettiFredenhagenVerch} entails the relations between
the algebras living on different manifolds. It is however a difficult
and not yet satisfactorily solved problem to find conditions which
single out physically meaningful states. All approaches to quantum
field theory on curved spacetime are confronted with this issue
(see for example \cite{Dappiaggi-2007}, \cite{Dappiaggi2009lecture}).\\
We thus consider asymptotically flat
space-time at timelike infinity that is to say a space-time with a
vanishing curvature at timelike infinity. We will always assume that
the curvature of the space-time is smooth enough and that it diminishes
sufficiently quickly when time goes to $\pm$infinity for the Fock space
to be properly defined (at least at timelike infinity). 

We then have two Fock spaces : one at time equals $-\infty$,
$\mathcal{F}_{in}$ and one at time equals $+\infty$, $\mathcal{F}_{out}$.
We want that the S-matrix simulates the evolution of a state of
$\mathcal{F}_{in}$ to $\mathcal{F}_{out}$ through the curved space-time.
One can find a more detailed description of this picture in \cite{Ford},
\cite{Wald}.

Unfortunately, linking the two Fock spaces is not an easy task for a general metric on the space-time. That's why we will restrict ourselves to a conformally flat metric which will provide us with simple equations and a trivial interpretation. A conformally flat metric is a metric $g'_{\mu\nu}$ such that :
\begin{align}
 g'_{\mu\nu}=\Omega^2(x)\eta_{\mu\nu}
\end{align}
with $\Omega$ a smooth positive function of $x$. Since we also want the space to be asymptotically flat, this means that :
\begin{align}
 \lim_{x_0\to -\infty} \Omega(x)=&A  & \lim_{x_0\to +\infty} \Omega(x)=&B~~~(A,~B\in\mathbb{R}_+^*)
\end{align}
And we will take $A=1$ to be able to interpret $\mathcal{F}_{in}$ as the usual free Fock space.

For a given metric $g_{\mu\nu}$ if one defines $g'_{\mu\nu}=\Omega^2(x)g_{\mu\nu}$, one has \cite{Dappiaggi2009lecture},\cite{Friedlander1975} :
\begin{align}
R'=&6\Omega^{-3}(x)\Box\Omega(x)+\Omega^{-2}(x)R
\\ (\Box'+\frac{1}{6}R')\varphi'(x)=&\Omega(x)^{-3}(\Box+\frac{1}{6}R)(\Omega(x) \varphi'(x))
\end{align}
with the primed objects related to $g'_{\mu\nu}$ and the unprimed ones
to $g_{\mu\nu}$. $R$ is the scalar curvature. This tells us that the
proper wave equation to be studied is $(\Box+\frac{1}{6}R)\varphi=0$
which, in the flat case, coincides with $\Box\varphi=0$. So in the
conformally flat case, one has :
\begin{align}
R'=&6\Omega^{-3}(x)\Box\Omega(x)
\\ (\Box'+\frac{1}{6}R')\varphi'(x)=&\Omega(x)^{-3}\Box(\Omega(x) \varphi'(x))=0
\end{align}
This means that one has :
\begin{align}
 \varphi'(x)=\Omega^{-1}(x)\varphi(x)=\frac{\Omega^{-1}(x)}{(2\pi)^{3/2}}\int{}d^3p~\frac{1}{\sqrt{2\omega_{\mathbf{p}}}}\left(e^{ipx}a_{\mathbf{p}}^{\dagger}+e^{-ipx}a_{\mathbf{p}}\right)
\end{align}
And by performing a change of variables, one has :
\begin{align}
  \varphi'(x)=\frac{1}{(2\pi)^{3/2}}\int{}d^3p~\frac{1}{\sqrt{2\omega_{\Omega^{-1/3}\mathbf{p}}}}\left(e^{i\frac{px}{\Omega^{1/3}}}a_{\mathbf{\frac{p}{\Omega^{1/3}}}}^{\dagger}+e^{-i\frac{px}{\Omega^{1/3}}}a_{\mathbf{\frac{p}{\Omega^{1/3}}}}\right)
\end{align}

By looking at this field, the interpretation is trivial. Indeed, one can
see that the positive and negative frequency parts are not mixed during
the evolution of the field in curved space-time. They don't ``see''
each other and so the conformally flat space-time doesn't make them
interact. And so the annihilation and creation operators of
$\mathcal{F}_{in}$ are respectively the same in $\mathcal{F}_{out}$
up to a change of the $|\vec{p}>_{in}\equiv |\vec{p}>$ basis to a
$|\vec{p}>_{out}\equiv|\vec{p'}>$ basis with $|\vec{p}>=|B^{1/3}\vec{p'}>$.
We can say here that $\mathcal{F}_{in}=\mathcal{F}_{out}$ since the
curvature of the conformal space doesn't really act on the state space
(at least not when one looks only at asymptotic limits).  And so, in
our case, one has :
\begin{align}
 |\vec{p_1},...,\vec{p_n}>\in\mathcal{F}_{in}\longrightarrow&|B^{1/3}\vec{p'_1},...,B^{1/3}\vec{p'_n}>\in\mathcal{F}_{out}=\mathcal{F}_{in}
\\ \mathrm{or}~~~  |\vec{q_1},...,\vec{q_n}>\in\mathcal{F}_{in}\longrightarrow&B^{-n}|B^{-1/3}\vec{q'_1},...,B^{-1/3}\vec{q'_n}>\in\mathcal{F}_{out}=\mathcal{F}_{in}
\end{align}
The S-matrix used in order to obtain this is the following (here is only the S-matrix for two particles going into two particles, the generalization being easy) :
\begin{align}
 S_{2,2,\mathrm{conf}}=&\frac{C_2}{2!}\left(\int{}d^3p_1d^3p_2{}~e^{2i\vec{Q}.\vec{p_1}(1-B^{-1/3})}a_{\mathbf{p_1}}^{\dagger}e^{i\vec{Q}.\left(\vec{p_1}(B^{-1/3}-1)+\vec{p_2}(1-B^{-1/3})\right)}a_{\mathbf{p_2}}^{\dagger}a_{\mathbf{p_1}}a_{\mathbf{p_2}}\right)\frac{C_2}{2!}
\end{align}
One has to remark that energy and momentum conservation still hold even though it is not directly apparent because of the change of basis. Moreover the S-matrix doesn't really simulate an evolution but is just the operator which maps the basis $|\vec{p}>\rightarrow |\vec{p'}>$. It is still interesting to see that the $|\vec{p}>$ basis and the $|\vec{q}>$ basis have the same transformation law under dilatation, only the dilatation factor is changed and they are the inverse of each other.

The formalism we have developed concerning the coordinate operators
permitted us to construct an S-matrix in curved space. Once again,
we have been able to do the calculations all the way through since
we have taken a simple case namely the conformally flat and
asymptotically flat case which turned out to be trivial (when
choosing the proper equation of motion).\\
Let us mention however, that also outside of our considerations
the problem of constructing such an S-matrix for a general metric
(and in particular for a general asymptotically flat space-time) is still open.

\subsection{Construction of fields and equations of motion}

In the free field case, we are given an action which gives us an equation of motion via Euler-Lagrange equation and this equation dictates us the form of the free field. In order to define new fields using the coordinate operators, we will use a somewhat different approach. Indeed, we will postulate a field $\varphi$ by using the known operators on the Fock space and assume that it is the expression of a $x$-dependent field at $x=0$. Then we will look at commutation relations involving this field and the energy-momentum operators. Finally, we will extend this field to a space-time by defining $\varphi(x)=e^{iPx}\varphi e^{-iPx}$.

In order to make things clear, let's use this method on the free field. We define a field:
\begin{align}
 \varphi=\int{}d^3p~\frac{1}{\sqrt{2\omega_{\mathbf{p}}}}(a_{\mathbf{p}}^{\dagger}+a_{\mathbf{p}})
\end{align}
with the fraction being a normalization and dimensional factor. Then we note that :
\begin{align}
 \Box_{P}(\varphi)\equiv-[P_{\mu},[P^{\mu},\varphi]]=0
\end{align}
And we define:
\begin{align}
\varphi(x)=e^{iPx}\varphi e^{-iPx}=\int{}d^3p~\frac{1}{\sqrt{2\omega_{\mathbf{p}}}}\left(e^{ipx}a_{\mathbf{p}}^{\dagger}+e^{-ipx}a_{\mathbf{p}}\right)
\end{align}
And since defining this $x$-dependent field gives $i[P_{\mu},\varphi(x)]=\partial_{\mu} \varphi(x)$, one has:
\begin{align}
 \Box_{x}(\varphi)\equiv-[P_{\mu},[P^{\mu},\varphi]]=0
\end{align}
that is to say, the field defined this way is a free field. Defining $\varphi(x)=e^{iPx}\varphi e^{-iPx}$ can be interpreted as being given a flat spacetime where the field $\varphi$ is a free field.

Let's now use the coordinate operators to define a field. We define:
\begin{align}\label{atilde}
\tilde{a}_{\mathbf{p}}=&e^{i\alpha p_0 N Q_0}a_{\mathbf{p}}e^{-i\alpha p_0 N Q_0}~(\alpha \in \mathbb{R})
\\ \varphi_1=&\int{}d^3p~\frac{1}{\sqrt{2\omega_{\mathbf{p}}}}(\tilde{a}_{\mathbf{p}}^{\dagger}+\tilde{a}_{\mathbf{p}})
\end{align}
And one has:
\begin {align}
[P_{\mu},\tilde{a}_{\mathbf{p}}^{\dagger}]=&(1-\alpha)p_{\mu}\tilde{a}_{\mathbf{p}}^{\dagger}
\\ [P_{\mu},\tilde{a}_{\mathbf{p}}]=&-(1-\alpha)p_{\mu}\tilde{a}_{\mathbf{p}}\label{commutatortilde}
\\ \Box_{P}(\varphi_1)\equiv &-[P_{\mu},[P^{\mu},\varphi_1]]=0
\\ \varphi_1(x)\equiv & e^{iPx}\varphi_1 e^{-iPx}=\int{}d^3p~\frac{1}{\sqrt{2\omega_{\mathbf{p}}}}\left(e^{i(1-\alpha)px}a_{\mathbf{p}}^{\dagger}+e^{-i(1-\alpha)px}a_{\mathbf{p}}\right) \label{xindu}
\\ \Box_{x}(\varphi_1(x))=&0 \label{xbox}
\end{align}

So a priori, this field is an ordinary free field in Minkowski space. But when looking more carefully at the commutator (\ref{xbox}), we see that instead of a standard Minkowski space-time, it could be a flat space-time with a metric proportional to $\eta_{\mu\nu}$. In order to show it, let's define coordinate operators in a massive case wich will permit us to see the effect of defining such a field instead of a standard free field.
So we define :
\begin{align}
\mathcal{L}=\frac{1}{2}\eta^{\mu \nu}\partial_{\mu} \varphi(x) \partial_{\nu} \varphi(x)-\frac{1}{2}m^2\varphi(x)^2
\end{align} 
which gives the following equation of motion :
\begin{align}
(\Box_x+m^2)\varphi(x)=0
\end{align} 
And when calculating $P_{\mu}$ one has :
\begin{align}
P_{\mu}^{(m)}=\int{}d^3p~p_{\mu}a_{\mathbf{p}}^{\dagger}a_{\mathbf{p}}
\end{align}
with $p_0=\omega_{\mathbf{p}}^{(m)}=\sqrt{-p^jp_j+m^2}$.
Then we define:

\begin{align}
X_0^{(m)}=&\int{}d^3p~\left((B_0-i)\frac{\omega_{\mathbf{p}}^{(m)}}{2\vec{p}^2}+(2A_0-B_0-i)\frac{1}{2\omega_{\mathbf{p}}^{(m)}}\right)a_{\mathbf{p}}^{\dagger}a_{\mathbf{p}}
-i\frac{\omega_{\mathbf{p}}^{(m)}}{\vec{p}^2}a_{\mathbf{p}}^{\dagger}p^l\frac{\partial a_{\mathbf{p}}}{\partial p^l} \nonumber\\ &\qquad\qquad (A_0,~B_0\in \mathbb{R})\nonumber
\\ Q_0^{(m)}=&X_0^{(m)} N^{-1}
\end{align}
Then one has:
\begin{align}
[P_0^{(m)},Q_0^{(m)}]=&i
\\ Q_0^{(0)}=&Q_0
\end{align}
So that if $m=0$ the definition of $Q_0^{(m)}$ is the same as in the massless case. Then one defines:
\begin{align}\label{atildem}
\tilde{a}_{\mathbf{p}}^{(m)}=&e^{i\alpha \frac{\vec{p}^2}{\omega_{\mathbf{p}}^{(m)}} N Q_0^{(m)}}a_{\mathbf{p}}e^{-i\alpha \frac{\vec{p}^2}{\omega_{\mathbf{p}}^{(m)}} N Q_0^{(m)}}~(\alpha \in \mathbb{R})
\\ \varphi_1^{(m)}=&\int{}d^3p~\frac{1}{\sqrt{2\omega_{\mathbf{p}}^{(m)}}}(\tilde{a}_{\mathbf{p}}^{(m)\dagger}+\tilde{a}_{\mathbf{p}}^{(m)})
\end{align}
which is still consistent with the massless case. And one has:

\begin{align}
[P_{\mu}^{(m)},\tilde{a}_{\mathbf{p}}^{(m)\dagger}]=&(1-\alpha)p_{\mu}^{(m)}\tilde{a}_{\mathbf{p}}^{(m)\dagger}
\\ [P_{\mu}^{(m)},\tilde{a}_{\mathbf{p}}^{(m)}]=&-(1-\alpha)p_{\mu}^{(m)}\tilde{a}_{\mathbf{p}}^{(m)}
\\ \Box_{P}(\varphi_1^{(m)})\equiv &-[P_{\mu}^{(m)},[P^{(m)\mu},\varphi_1^{(m)}]]=-(1-\alpha)^2m^2\varphi_1^{(m)}
\\ \varphi_1^{(m)}(x)\equiv & e^{iP^{(m)}x}\varphi_1^{(m)} e^{-iP^{(m)}x}\nonumber\\
=&\int{}d^3p~\frac{1}{\sqrt{2\omega_{\mathbf{p}}}}\left(e^{i(1-\alpha)p^{(m)}x}a_{\mathbf{p}}^{(m)\dagger}+e^{-i(1-\alpha)p^{(m)}x}a_{\mathbf{p}}^{(m)}\right) \label{xindum}
\\ \Box_{x}(\varphi_1(x)^{(m)})=&-(1-\alpha)^2m^2\varphi_1^{(m)} \label{xboxm}
\\ \frac{1}{(1-\alpha)^2}\Box_{x}(\varphi_1(x)^{(m)})=&-m^2\varphi_1^{(m)}
\end{align}
When looking at the last two equations, we understand why we couldn't see the effect of redefining $a_{\mathbf{p}}$ in the massless case: if $m=0$, one can't see the factor $(1-\alpha)^2$. We have then two possible interpretations. Either one can say that $\varphi_1^{(m)}(x)$ is a free massive field of mass $(1-\alpha)m$ in Minkowski space-time. Or one can say that $\varphi_1^{(m)}(x)$ is a free massive field of mass $m$ in a flat spacetime with metric $\frac{1}{(1-\alpha)^2}\eta_{\mu\nu}$. So we have been able, starting with the standard free case, and only introducing coordinate operators, to have a control on the space-time by redefining the objects considered namely a field.\\
Another point can be made when starting from the result. For $\alpha \ne 1$ the
operators $\tilde a^\dagger$ and $\tilde a$ are effectively nothing but creation, resp.\
annihilation operators obtained from $a^\dagger$, resp. $a$ by a {\sl finite}
dilatation including a scaling of the mass. For $\alpha=1$ however the
``dilatation''
counterbalances exactly the translation, the field does
not ``see'' a spacetime 
and does not evolve in it. This originates from non-locality of $X_0$
on the one hand and of the specific
choice of the function of $\vec{p}$ in the exponential in (\ref{atilde}),
(\ref{atildem}) on the other
hand. ($X_0$, indeed, since the dependence of $N$ drops out.)\\
Certainly a definition based only on the fundamental operators $P_\mu$
and $Q_\mu$, e.g.\ $i\alpha P^{\mu}Q_{\mu}$ in the exponent, would be more
natural in the present context, however technically much more involved.
Still, we consider the example to be instructive.\\

\section{Discussion, conclusions, open questions}

We studied operators $Q_\mu$ conjugate to the energy-momentum operators
$P_\mu$ within 
quantum field theory of one scalar field. Searching in Fock space and 
limiting the search to the three-dimensional integral of bilinears of 
$a^\dagger$ and $a$  multiplied by factors $p_\mu$ and derivatives 
w.r.t.\ $p_j$ such that appropriate dimensions result, we first found out
that demanding locality in field space singles out $K_\mu$, the generators
of conformal transformations. This we called the ``conformal case'' to be 
treated in \cite{SiboldEden}. {\sl Generic} is thus the {\sl non-local}
form of $Q_\m$, 
we named this the ``basic'' case, being pursued here. Amongst the huge 
number of possible $Q_\mu$ there are no Lorentz covariant ones; with
necessity non-vanishing is the commutator 
$\lbrack Q_0,Q_j \rbrack$, hence non-trivial mixing of time and space
coordinates. We find ourselves thus in the context of non-commutative geometry
and do not worry about the lack of Lorentz covariance because the latter
happens also to appear in a Moyal deformed quantum field theory. Here
the situation is even more favorable since one might be able to construct
physical quantities as functions of the $Q_\mu$ which are covariant.
We further selected according to computational simplicity ({\sl commuting} space/space coordinates).\\
The eigenvalue problems of $X_\mu$, resp.\ $Q_\mu$ can be solved easily;
we further assume the eigenvalues to be real. This is a non-trivial
assumption,
pointing to the fact that the unbounded operators involved, are, to begin
with, only Hermitian and not necessarily self-adjoint. As a first step to describe their domains
we confirm that the conjugation commutators hold in all bases which we used
explicitly: the basis $|\vec{p}>$ of vectors in Fock space; the basis $|q_j>$ spanned by the eigenvectors of $Q_j$ and finally the basis $|q_0>$ spanned
by the eigenvectors of $Q_0$. (Recall: $Q_0$ and $Q_j$ do not commute.)
It is interesting to observe that we obtain a continuum normalization
within the $|q_0>$ states only after having formed time reversal even/ odd
combinations of them. We understand this as a signal how to escape
the consequences of Paulis theorem \cite{Pauli} because it enforces
the restriction
of $q_0$ to the non-negative reals.\\

These are the technical preliminaries. What about physical consequences?
We elaborate on them in two examples: the $S$-matrix and fields 
$\phi(Q)$. Using the operators $Q_j$ we prescribe explicitly matrix elements
of the $S$-matrix from 2-particle-in-states to 2-particle-out-states
(generalizations are obvious). By construction the $S$-matrix is unitary.
The interpretation of this $S$-matrix as an evolution operator from a 
Fock space $\mathcal{F}_{in}$ at timelike $-$infinity to a Fock space 
$\mathcal{F}_{out}$ at timelike $+$infinity through a curved spacetime
is in 
general highly non-trivial, hence we assume conformal and asymptotic
flatness for the underlying spacetime. This simplified case can then
be mimicked easily by choosing appropriately the general functions employed
in the prescription for $S$. And the $S$-operator amounts to a change of basis
in the in/out-Fock space. Although this example is very simple it nevertheless
shows that our general machinery works.\\
In analogy to the free scalar field in Minkowski spacetime we prescribe 
$Q_0$-transformed creation and annihilation operators combined to form a field
which satisfies (\ref{xbox}), seemingly a free field equation. When enlarging
the framework to comprise a non-vanishing mass, it turns out however that
the resulting free field equation (\ref{xboxm}) can be interpreted as one of a scalar field
with a different mass or as a free scalar field with the same mass on a Minkowski
space with a different metric. This result clearly shows that the formalism
developped here is suited to describe changes in the structure of spacetime
as arising from coordinate operators.\\

Let us now put our results into perspective of present day theory. 
Obviously one can expect non-trivial uncertainty relations between $Q_0$ and $Q_j$,
once one has chosen a suitable set of states. This is to be compared with
the discussion in \cite{DoplicherFredenhagenRoberts}, where their role is
explained in detail.
As the interesting paper \cite{BorrisVerch} shows there is an overlap of methods
and techniques used to formulate problems of quantum field theory on curved
spacetime and quantum field theory based on noncommutative geometry. The same
is true here: the non-trivial algebra of the $Q_\mu$ yields at once a non-trivial
$S$-matrix with information on a potential underlying spacetime as it permits
to define fields as functions of coordinate operators which eventually have
to be evaluated in Fock space with a spacetime to be found. For the latter
we used indeed via (\ref{xindu}), (\ref{xindum}) conventional spacetime (induced representation)
as a probe. Here a more intrinsic procedure has to be found. This is also true
for the interpretation of the $S$-matrix: we first made an assumption on the underlying spacetime and could then produce this by choosing the free functions 
in the $S$-matrix. Clearly the converse is to be looked for: for a given function
in $S$ we would like to see which spacetime is associated with it. This problem
is however to be faced also in other approaches \cite{Dappiaggi-2007}, \cite{Dappiaggi2009lecture}. It is the inverse scattering problem: deducing from 
scattering data underlying structures.\\
There are other open questions. Mathematical ones: Which are the domains of our operators?
Which is the relation of
the present approach to the characterization of e.g.\ Poisson structures as found
by Kontsevich \cite{Kontsevich}? Again, physical ones:  Could one generalize from the outset
to the interacting theory? Could one expect at some level effectively
Lorentz covariance and causality?\\

\noindent
Acknowledgements: One of us (G.S.) would like to thank the Institut
f\"ur Theoretische Physik for hospitality extended to him.
K.S.\ thanks B.\ Eden for helpful discussions. The enlightening
remarks of R.\ Verch to the more mathematical literature are most
gratefully acknowledged. We also thank the referee for clarifying remarks
and for pointing out shortcomings of our presentation in a very constructive
fashion.\\

\section{Appendix: Notations, conventions, useful formulae}
We work with a scalar field, given in terms of creation and annihilation operators ($a_\mathbf{p}^{\dagger}$ and $a_\mathbf{p}$ respectively):
\begin{align}
\varphi(x)=&\frac{1}{(2\pi)^{3/2}}\int{}d^3p~\frac{1}{\sqrt{2\omega_{\mathbf{p}}}}\left(e^{ipx}a_{\mathbf{p}}^{\dagger}+e^{-ipx}a_{\mathbf{p}}\right)\label{freefield}
\\ a_{\mathbf{p}}=&\frac{1}{(2\pi)^{3/2}}\int{}d^3x~\frac{i}{\sqrt{2\omega_{\mathbf{p}}}}\left(-\partial_{x_0}(e^{ipx})\varphi(x)+e^{ipx}\partial_{x_0}\varphi(x)\right)
\\ [a_{\mathbf{p}},a_{\mathbf{p'}}^{\dagger}]=&\delta(\vec{p}-\vec{p'})
\end{align}
with $\omega_{\mathbf{p}}=\sqrt{\vec{p}^2+m^2}=p_0$. Most often we deal with the massless case, $m=0$. A basis of the Fock space is spanned by the momentum eigenstates $|\vec{p_1},\vec{p_2},....,\vec{p_n}>$ defined by the action of $a_{\mathbf{p_1}}^{\dagger}a_{\mathbf{p_2}}^{\dagger}...a_{\mathbf{p_n}}^{\dagger}$ on the vacuum such that $P_{\mu}|\vec{p_1},\vec{p_2},....,\vec{p_n}>=(p_{1\mu}+p_{2\mu}+...+p_{n\mu})|\vec{p_1},\vec{p_2},....,\vec{p_n}>$.

3-dimensional Fourier transform :
\begin{align}
 \tilde{f}(\vec{q})=FT(f(\vec{p}))=\frac{1}{(2\pi)^{3/2}}\int{}d^3pe^{i\vec{q}.\vec{p}}f(p)\nonumber
\\ f(\vec{p})=FT^{-1}(\tilde{f}(\vec{q}))=\frac{1}{(2\pi)^{3/2}}\int{}d^3qe^{-i\vec{q}.\vec{p}}\tilde{f}(q)\nonumber
\end{align}

Conformal transformations:\\
$\bullet$ currents:\\
\begin{itemize}
\renewcommand{\labelitemi}{$-$}
\item the improved energy-momentum tensor ($m=0$):
\begin{equation}
 T_{\mu \nu}=\partial_{\mu}\varphi\partial_{\nu}\varphi-\frac{1}{2}\eta_{\mu \nu}\partial^{\rho}\varphi\partial_{\rho}\varphi-\frac{1}{4}\eta_{\mu \nu}\varphi\Box\varphi-\frac{1}{6}(\partial_{\mu}\partial_{\nu}-\eta_{\mu \nu}\Box)\varphi^2
\end{equation}
\item the Lorentz current : $M_{\mu \nu \rho}=x_{\mu}T_{\nu \rho}-x_{\nu}T_{\mu \rho}$
\item the dilatation current : $D_{\mu}=x^{\nu}T_{\mu \nu}$
\item the conformal current : $K_{\rho \mu}=(2x_{\rho}x^{\lambda}-\eta_{\rho}^{\lambda}x^2)T_{\lambda \mu}$
\end{itemize}
$\bullet$ charges (generators of the respective transformation):\\
\begin{align}
P_{\mu}=&\int{}d^3x~T_{\mu 0}=\int{}d^3p~p_{\mu}a_{\mathbf{p}}^{\dagger}a_{\mathbf{p}}
\\ M_{\mu \nu}=&\int{}d^3x~(x_{\mu}T_{\nu 0}-x_{\nu}T_{\mu 0})
\\ M_{j0}=&\frac{i}{2}\int{}d^3p~p_0\left((\frac{\partial}{\partial p_j}a_{\mathbf{p}}^{\dagger})a_{\mathbf{p}}-a_{\mathbf{p}}^{\dagger}(\frac{\partial}{\partial p_j}a_{\mathbf{p}})\right)=i\int{}d^3p~\frac{p_j}{2p_0}a_{\mathbf{p}}^{\dagger}a_{\mathbf{p}}-p_0a_{\mathbf{p}}^{\dagger}\frac{\partial}{\partial p^j}a_{\mathbf{p}}
\\M_{jk}=&\frac{i}{2}\int{}d^3p\left(p_k\left((\frac{\partial}{\partial p^j}a_{\mathbf{p}}^{\dagger})a_{\mathbf{p}}-a_{\mathbf{p}}^{\dagger}(\frac{\partial}{\partial p^j}a_{\mathbf{p}})\right)-p_j\left((\frac{\partial}{\partial p^k}a_{\mathbf{p}}^{\dagger})a_{\mathbf{p}}-a_{\mathbf{p}}^{\dagger}(\frac{\partial}{\partial p^k}a_{\mathbf{p}})\right)\right)\nonumber
\\ M_{jk}=&i\int{}d^3p~p_ja_{\mathbf{p}}^{\dagger}\frac{\partial}{\partial p^k}a_{\mathbf{p}}-a_{\mathbf{p}}^{\dagger}p_k\frac{\partial}{\partial p^j}a_{\mathbf{p}}
\\D=&\int{}d^3x~x^{\lambda}T_{\lambda 0}
\\ D=&\frac{i}{2}\int{}d^3p~p^l\left((\frac{\partial}{\partial p^l}a_{\mathbf{p}}^{\dagger})a_{\mathbf{p}}-a_{\mathbf{p}}^{\dagger}(\frac{\partial}{\partial p^l}a_{\mathbf{p}})\right)=-i\int{}d^3p~\frac{3}{2}a_{\mathbf{p}}^{\dagger}a_{\mathbf{p}}+a_{\mathbf{p}}^{\dagger}p^l\frac{\partial}{\partial p^l}a_{\mathbf{p}}
\\ K_{\mu}=&\int{}d^3x(2x_{\mu}x^{\nu}-\eta_{\mu}^{\nu}x^2)T_{\nu 0}
\\ K_0=&\int{}d^3p~\frac{1}{2}p_0\left((\frac{\partial}{\partial p^l}\frac{\partial}{\partial p_l}a_{\mathbf{p}}^{\dagger})a_{\mathbf{p}}+a_{\mathbf{p}}^{\dagger}(\frac{\partial}{\partial p^l}\frac{\partial}{\partial p_l}a_{\mathbf{p}})\right)+\frac{1}{4}\frac{1}{p_0}a_{\mathbf{p}}^{\dagger}a_{\mathbf{p}}\nonumber
\end{align}
\begin{align}
 K_0=&\int{}d^3p~-\frac{3}{4}\frac{1}{p_0}a_{\mathbf{p}}^{\dagger}a_{\mathbf{p}}-\frac{1}{p_0}a_{\mathbf{p}}^{\dagger}p^l\frac{\partial}{\partial p^l}a_{\mathbf{p}}+p_0a_{\mathbf{p}}^{\dagger}\frac{\partial}{\partial p^l}\frac{\partial}{\partial p_l}a_{\mathbf{p}}
\\ K_j=&\int{}d^3p~-\frac{1}{4}\frac{p_j}{p_0^2}a_{\mathbf{p}}^{\dagger}a_{\mathbf{p}}+(\frac{\partial}{\partial p_j}a_{\mathbf{p}}^{\dagger})(p^l\frac{\partial}{\partial p^l}a_{\mathbf{p}})+(p^l\frac{\partial}{\partial p^l}a_{\mathbf{p}}^{\dagger})(\frac{\partial}{\partial p^j}a_{\mathbf{p}})-p_j\frac{\partial}{\partial p^l}a_{\mathbf{p}}^{\dagger}\frac{\partial}{\partial p_l}a_{\mathbf{p}}\nonumber
\\ K_j=&\int{}d^3p~-\frac{1}{4}\frac{p_j}{p_0^2}a_{\mathbf{p}}^{\dagger}a_{\mathbf{p}}-3a_{\mathbf{p}}^{\dagger}\frac{\partial}{\partial p^j}a_{\mathbf{p}}-2a_{\mathbf{p}}^{\dagger}p^l\frac{\partial}{\partial p^l}\frac{\partial}{\partial p^j}a_{\mathbf{p}}+p_ja_{\mathbf{p}}^{\dagger}\frac{\partial}{\partial p^l}\frac{\partial}{\partial p_l}a_{\mathbf{p}}
\end{align}

All these operators are Hermitian and local (i.e.\ they have a local expression in term of fields). And the commutation relations between them define the conformal algebra which is closed :
\begin{align}
[P_{\mu},P_{\nu}]=&0 & [P_0,M_{l0}]=&-iP_l & [P_0,M_{lk}]=&0 \nonumber
\\ [P_j,M_{l0}]=&i \delta_{jl} P_0 \hspace{2cm}& [P_j,M_{lk}]=&i \delta_{jl} P_k-i\delta_{jk} P_l \hspace{2cm}& [P_{\mu},D]=&iP_{\mu}\nonumber
\\ [P_{\mu},K_{\nu}]=&2i(\eta_{\mu\nu}D-M_{\mu\nu})& [D,M_{j0}]=&0 & [D,M_{lk}]=&0 \nonumber
\\ [D,K_{\mu}]=& iK_{\mu} & [M_{j0},M_{lk}]=&i\delta_{jl}M_{k0}-i\delta_{jk}M_{l0} & [M_{jk},M_{jm}]=&iM_{km}\nonumber
\\ [M_{j0},K_0]=&iK_j & [M_{jk},K_0]=&0 & [M_{j0},K_l]=&-i\delta_{jl}K_0\nonumber
\\ [M_{jl},K_m]=&i\delta_{lm}K_j-i\delta_{jm}K_l & [K_{\mu},K_{\nu}]=&0\nonumber
\end{align}

\noindent
The expressions of the charges in terms of
creation and annihilation operators have been worked out in
collaboration with B.\ Eden which is gratefully acknowledged.\\

 
\providecommand{\href}[2]{#2}\begingroup\raggedright\endgroup

\end{document}